\documentclass[preprint,showpacs,amsmath,amssymb,aps,prd]{revtex4}
\usepackage{graphicx}

\usepackage{amsmath}

 \newcommand{\del}
{\partial}
 \newcommand{\cM}{{\cal M}}

 \newcommand{\ds}{\displaystyle}
 
 \newcommand{\diag}{\mathop{\rm
diag}\nolimits} \newcommand{\orto}{{\scriptscriptstyle\perp}}
  \newcommand{\Pn}
{ {P_{\orto}}\vphantom{P} }

\begin{document}

\title{Pseudo-Riemannian Universe from Euclidean bulk}

\author{Milovan Vasili\'c} \email{mvasilic@ipb.ac.rs}
\affiliation{Institute of Physics, University of Belgrade, P.O.Box 57,
11001 Belgrade, Serbia}

\date{\today}

\begin{abstract}
I develop the idea that our world is a brane-like object embedded in
Euclidean bulk. In its ground state, the brane constituent matter is
assumed to be homogeneous and isotropic, and of negligible influence
on the bulk geometry. The analysis of this paper is model
independent, in the sense that action functional of bulk fields is
not specified. Instead, the behavior of the brane is derived from the
universally valid conservation equation of the bulk stress tensor.
The present work studies the behavior of a $3$-sphere in the
$5$-dimensional Euclidean bulk. The sphere is made of bulk matter
characterized by the equation of state $p=\alpha\rho$. It is shown
that stability of brane vibrations requires $\alpha < 0$. Then, the
stable brane perturbations obey Klein-Gordon-like equation with an
effective metric of Minkowski signature. The argument is given that
it is this effective metric that is detected in physical
measurements. The corresponding effective Universe is analyzed for
all the values of $\alpha<0$. In particular, the effective metric is
shown to be a solution of Einstein's equations coupled to an
effective perfect fluid. The effective energy density and pressure at
the present epoch are calculated. So are the age of the Universe, and
the effective cosmological constant. All the results are presented in
two tables. As an illustration, one simple choice of the brane
constituent matter is studied in detail.
\end{abstract}

\pacs{04.50.-h, 11.27.+d}

\maketitle

\section{Introduction}\label{Sec1}

The mainstream braneworld physics, which has extensively been
developed after the seminal work of Randall and Sundrum
\cite{RS1,RS2}, typically employs pseudo-Riemannian bulks. In this
paper, I am investigating the idea that our world is a domain wall
embedded in a higher dimensional Euclidean space. This is motivated
by the observation that Euclidean bulk may improve stability
properties of the brane Universe. Indeed, it is straightforwardly
verified that spherical branes embedded in Minkowski bulk are
unstable, irrespectively of the nature of the brane constituent
matter. In particular, the dynamics of Nambu-Goto $3$-sphere is
burdened with the presence of tachyons.

The general idea behind this line of research is an attempt to obtain
the observed physics as a physics of brane vibrations. This is a kind
of geometrization program, in which observable matter is identified
with disturbances of the braneworld vacuum solution. In this respect,
it resembles the string theory. The brane vibrations are expected to
model physical fields the same way as string vibrations model
elementary particles.

Before I continue, let me explain why I believe this geometrization
program may work. I start with the observation that our physical
fields can be seen as a $4$-dimensional surface embedded in a
higher-dimensional bulk. Indeed, the parametric form of a $4$-surface
in a $D$-dimensional bulk is defined by $D$ functions of $4$
coordinates. If the $4$-surface is interpreted as the trajectory of a
$3$-brane, the field equations become the brane world-sheet
equations. This way, the behavior of physical fields becomes the
behavior of a $3$-brane. In a perturbative analysis, the physical
fields are associated with vibrations of the brane vacuum. Then, the
reparametrization invariant world-sheet equations describe $D-4$
field components. This way, every conventional field theory can be
rewritten in terms of brane vibrations. An early attempt along these
lines is the construction of geodetic brane gravity
\cite{b1,b2,b3,b4}. With this thought in mind, I choose the simplest
possible scenario to illustrate this idea.

In this paper, I shall study small perturbations of a simple
braneworld vacuum. I begin with the assumption that the brane is made
of matter fields that live in a $5$-dimensional Euclidean bulk. The
analysis is {\it model independent}, in the sense that action
functional of the bulk matter is not specified. Instead, I use the
universally valid conservation equations of the stress tensor. The
problem with this approach is that the brane interior may have its
own degrees of freedom. This may compromise the original idea that
observable matter stems from the brane vibrations alone. To prevent
this, I shall assume that the brane interior is hard to excite. This
can always be achieved by a proper choice of the bulk matter. The
needed ground state of the brane is searched for in the form of a
homogeneous and isotropic $3$-sphere. This way, my program reduces to
the construction of a braneworld Universe. (The idea to describe
Universe as a spherical shell has first been considered in
\cite{a1,a2,a3}, and more recently in \cite{a4,a5}.) Let me summarize
the basic assumptions of this program:
\begin{itemize}
\item The arena for my considerations is a higher-dimensional space
called bulk.

\item The dynamics of bulk fields is governed by a diffeomorphism
invariant action functional.

\item Diffeomorphism invariance implies the existence of the
conserved stress tensor.

\item The bulk action is assumed to possess a brane-like
kink solution. Such a kink resembles a 4-dimensional surface
called world-sheet.

\item The presented analysis is model independent, in the sense that
the action is not explicitly specified. Instead, the
world-sheet equations are derived from the universally valid
conservation equations of the bulk stress tensor.

\item The brane is assumed to have negligible influence on the bulk
geometry. Consequently, the bulk geometry is well
approximated by the flat Euclidean space.

\item For simplicity, the considerations of this work are restricted
to the 5-dimensional bulk.
\end{itemize}
I start with the conservation equation of the bulk stress tensor,
\begin{equation} \label{1}
\nabla_{\nu}T^{\mu\nu}=0 \,,
\end{equation}
and apply it to bulk fields that form a brane-like kink configuration
called world-sheet. Throughout the paper, it is assumed that the brane
has negligible influence on the bulk geometry. In other words, I work
in the probe matter approximation. (An early model of probe
brane-worlds has been proposed in \cite{b1}, and later developed in
\cite{b2,b3,b4}.) The derivation of the manifestly covariant
world-sheet equations is developed in \cite{VV1,VV2}, as a
generalization of the Mathisson-Papapetrou method for point-like
matter \cite{M,P}. According to this, the stress tensor
$T^{\mu\nu}(x)$, which is well localized around the
$(p+1)$-dimensional surface $\cM$ in a $D$-dimensional bulk, has an
expansion as a series of $\delta$-function derivatives. For infinitely
thin branes of Euclidean nature, this multipole expansion reduces to
\begin{equation} \label{2}
T^{\mu\nu} = \int_{\cM} d^{p+1}\xi \sqrt{\gamma}\,
M^{\mu\nu} \frac{\delta^{(D)}(x-z)}{\sqrt{g}} \,.
\end{equation}
The surface $\cM$ is defined by the equation $x^{\mu}=z^{\mu}(\xi)$,
where $\xi^a$ are the surface coordinates, and $M^{\mu\nu}(\xi)$ are
free coefficients. The surface coordinate vectors $u_a^{\mu} \equiv
\del z^{\mu}/\del\xi^a$ define the surface induced metric
$\gamma_{ab}\equiv g_{\mu\nu}u_a^{\mu}u_b^{\nu}$. The bulk
metric is denoted by $g_{\mu\nu}(x)$.

The decomposition (\ref{2}) is used as an ansatz for solving the
conservation equation (\ref{1}). This has already been done in
\cite{VV2}, where manifestly covariant $p$-brane world-sheet equations
and boundary conditions have been derived. These equations determine
the behavior of the embedding functions $z^{\mu}(\xi)$, and the
coefficients $M^{\mu\nu}(\xi)$. Skipping the details of the
calculation, I shall display the final result. The equation (\ref{1})
leads to the world-sheet equations
$$
\Pn^{\mu}_{\lambda}M^{\nu\lambda} = 0 \,, \qquad
\nabla_a\left(M^{\mu\nu}u^a_{\nu}\right) = 0 \,,
$$
where $\Pn^{\mu}_{\nu} \equiv \delta^{\mu}_{\nu} -
u_a^{\mu}u^a_{\nu}$ is the orthogonal projector to the world-sheet,
and $\nabla_a$ stands for the total covariant derivative that acts on
both types of indexes. The boundary conditions are trivially
satisfied. The equation $\Pn^{\mu}_{\lambda} M^{\nu\lambda} = 0$
tells us that $M^{\mu\nu}$ coefficients lack the orthogonal
components. Therefore, their general form is
$$
M^{\mu\nu} = m^{ab} u^{\mu}_a u^{\nu}_b \,,
$$
where $m^{ab}(\xi)$ are the residual free coefficients. In terms of
$m^{ab}$, the brane world-sheet equations are rewritten as
\begin{equation} \label{3}
\nabla_a\left(m^{ab}u^{\mu}_b\right) = 0 \,.
\end{equation}
It is straightforward to verify that the equations (\ref{3}) imply the
covariant conservation law
$$
\nabla_a m^{ab} = 0 \,.
$$
For this reason, the coefficients $m^{ab}$ will be referred to as the
brane stress tensor.

The results obtained in this paper are summarized as follows. (a) The
ground state stress tensor $m^{ab}$ is characterized by the generic
equation of state $p=p(\rho)$, independently of the nature of the
brane constituent matter. In the sector of small $\rho$, this
equation takes the widely exploited form $p=\alpha\rho$. (b) The
stability analysis is presented. I demonstrate that the condition
$\alpha<0$ is a universal necessary condition to ensure stable ground
state of the brane Universe. With this condition satisfied, the
infinitesimal brane vibrations obey Klein-Gordon-like equation with
an effective metric of Minkowski signature. If these vibrations are
identified with the observable matter, the obtained effective metric
defines the observed pseudo-Riemannian geometry. The corresponding
effective Universe is investigated for a range of values of
$\alpha<0$. It is shown that the effective metric, coupled to an
effective perfect fluid, becomes a solution of Einstein's equations.
This way, the effective Universe is rewritten in terms of the
standard general relativistic (GR) cosmology. The present time values
of the effective energy density and pressure are calculated along
with the effective cosmological constant and the age of the Universe.
The results are presented in two tables. (c) As an example, the brane
made of an incompressible fluid is studied in detail. The stability
of both, brane vibrations and brane interior, is proven for all
$\alpha\leq -1/3$. (d) It is demonstrated that bulk conservation
equations do not allow braneworlds of finite lifetime. In particular,
the Big Bang cosmology is not supported by the typical braneworld
concept. I ague, however, that these solutions are perfectly
legitimate if restricted to sectors which are not close to
singularities.

The conventions used throughout the paper are the following. Greek
indices $\mu,\nu,\dots$ are the bulk indices, and run over
$0,1,2,3,4$. Latin indices $a,b,\dots$ are the braneworld indices,
and run over $0,1,2,3$. The coordinates of the bulk and brane are
denoted by $x^{\mu}$ and $\xi^a$, respectively. The bulk metric is
the flat Euclidean metric $\delta_{\mu\nu}$, and the induced
braneworld metric is $\gamma_{ab}(\xi)$. The signature convention is
defined by $\diag(+,+,\dots,+)$, and the indices are raised using the
inverse metrics $\delta^{\mu\nu}$ and $\gamma^{ab}$.

\section{Ground state solution}\label{Sec2}

Supported by astronomical observations, the ground state of the
Universe is conventionally considered spatially homogeneous and
isotropic. This leaves us with a limited choice of ground state
branes and allowed stress-energy tensors. In this paper, the ground
state brane is assumed to be a $3$-sphere embedded in the
$5$-dimensional Euclidean space. Its world-sheet
$x^{\mu}=z^{\mu}(\xi)$ is defined by
\begin{eqnarray}
& z^0=t,\ z^1=\ell(t)\cos\psi,\
z^2=\ell(t)\sin\psi\cos\theta, & \nonumber \\ &
z^3=\ell(t)\sin\psi\sin\theta\cos\phi \,, & \label{4} \\ &
z^4=\ell(t)\sin\psi\sin\theta\sin\phi \,, & \nonumber
\end{eqnarray}
where the identification $\xi^0\equiv t$, $\xi^1\equiv\psi$,
$\xi^2\equiv \theta$, $\xi^3\equiv \phi$ is used. The $t$ coordinate
takes values on the whole real axis, and $\psi, \theta, \phi$ are
the standard angular coordinates of the $3$-sphere ($\psi\in[0,\pi]$,
$\theta\in[0,\pi]$, $\phi\in[0,2\pi]$). Then, one readily calculates
the coordinate vectors $u^{\mu}_a \equiv \del z^{\mu}/\del \xi^a$,
and the induced metric $\gamma_{ab} \equiv \delta_{\mu\nu} u^{\mu}_a
u^{\nu}_b$. The induced metric is read from
\begin{equation} \label{5}
ds^2 = \big(\dot\ell^2 + 1 \big)
dt^2 + \ell^2 \left[ d\psi^2 + \sin^2\psi \left( d\theta^2 +
\sin^2\theta\, d\phi^2 \right) \right] ,
\end{equation}
where $\dot\ell \equiv d\ell/dt$. The radius $\ell(t)$ is an
undetermined variable of the brane equations (\ref{3}).

The general form of the homogeneous and isotropic world-sheet stress
tensor $m^{ab}$ is
\begin{equation}\label{6}
m^{ab} = \left(
\begin{array}{cc}
\rho\gamma^{00} &                    0                 \\
           0                & p\gamma^{\alpha\beta} \\
\end{array}\right) ,
\end{equation}
where $\alpha, \beta = 1, 2, 3$, and the variables $\rho = \rho(t)$
and $p=p(t)$ depend on $t$ only. It is easily checked that only two
out of five equations (\ref{3}) are independent, which is
insufficient to determine the three unknown variables $\rho$, $p$ and
$\ell$. One needs to further specify the nature of the brane
constituent matter. In what follows, I shall demonstrate that the
equation of state $p=\alpha\rho$, which is conventionally imposed by
hand, is in fact, universal in the sector of small $\rho$. Indeed,
the assumption of spatial homogeneity ensures that $\rho = \rho(t)$
and $p=p(t)$, which is nothing but the parametric form of the
relation $p=p(\rho)$. When expanded in a power series around
$\rho=0$, it takes the form
$$
p = \alpha_0 + \alpha_1\rho + \alpha_2\rho^2 + \cdots \,,
$$
which reduces to $p=\alpha_0+\alpha_1\rho$ in the sector of small
$\rho$. In the conventional physics, the condition $\rho = 0$ implies
$p=0$, and therefore, $\alpha_0 = 0$. Although my considerations in
this section are not conventional, it will be shown later that they
lead to a conventional effective Universe. For this reason, the
condition $\alpha_0=0$ is needed in my considerations, too. Thus, I am
left with the approximate equation of state
\begin{equation} \label{7}
p=\alpha\rho
\end{equation}
{\it applicable in the sector of small $\rho$}. Exotic equations of
state, which can not be expanded in a power series around $\rho =0$,
can still be approximated by (\ref{7}) in the close vicinity of $\rho
=0$. Indeed, in the limit $\rho\to 0$, $\alpha\to\infty$, the exotic
equations of state are practically indistinguishable from (\ref{7}).

The world-sheet equations (\ref{3}) with the equation of state
(\ref{7}) are straightforwardly solved. As it turns out,
only two out of five equations are independent. The equation
$$
\frac{\dot\rho}{\rho} = 3\left(\alpha-1\right)\frac{\dot\ell}{\ell}
$$
follows from the conservation equation of the brane stress tensor
$m^{ab}$, while
$$
\frac{d}{dt}\left(\frac{\ell^{3\alpha}}{\ds\sqrt{1+\dot\ell^2}}\right)=0
$$
is derived from the zero component of the world-sheet equations
(\ref{3}). In what follows, the coordinate $t$ will be replaced by
$\eta$, as defined by
\begin{equation} \label{8}
\ds \sqrt{1+\dot\ell^2}\, dt = \ell d\eta \,.
\end{equation}
In terms of $\eta$, the induced metric takes the form
\begin{equation} \label{9}
ds^2 = R^2(\eta)\left[d\eta^2 + d\psi^2 +
\sin^2\psi \left( d\theta^2 + \sin^2\theta\, d\phi^2 \right) \right] ,
\end{equation}
and the solution for $R(\eta)\equiv \ell(t(\eta))$ reads
\begin{subequations} \label{11}
\begin{equation}\label{11a}
R = R_0\exp\left(\frac{\eta}{\eta_0}\right) \quad
{\rm for} \quad \alpha=0\,,
\end{equation}
\begin{equation} \label{11b}
R = R_0\big[\cosh(3\alpha\eta)\big]^{\frac{1}{3\alpha}} \quad
{\rm for}\quad\alpha\neq 0\,.
\end{equation}
\end{subequations}
The solution for $\rho\equiv \rho(t(\eta))$ is
\begin{equation} \label{10}
\rho = \rho_0
\left(\frac{R}{R_0}\right)^{3\left(\alpha-1\right)}\,.
\end{equation}
The parameters $\rho_0$, $R_0$ and $\eta_0$ are integration constants.

\begin{figure}[htb]
\begin{center}
\includegraphics[height=5cm]{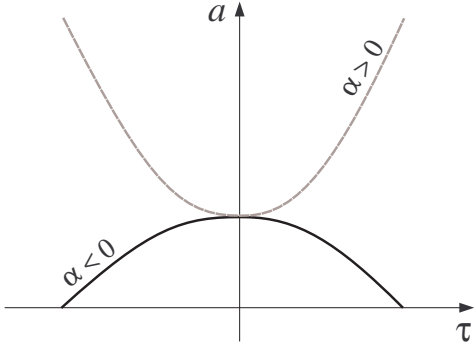}
\end{center}
\vspace*{-.5cm}
\caption{Behavior of the brane Universe.\label{f1} }
\end{figure}

The behavior of the $3$-sphere can also be parametrized by the
proper parameter $\tau$. It is defined by $d\tau = R\, d\eta$ whereupon
the induced metric takes the form
$$
ds^2 = d\tau^2 + a^2 \left[ d\psi^2 +
\sin^2\psi \left( d\theta^2 + \sin^2\theta\, d\phi^2 \right) \right] .
$$
The $\tau$ dependence of the radius $a(\tau) \equiv R(\eta(\tau))$
for $\alpha > 0$ and $\alpha < 0$ is shown in Fig. \ref{f1}. For
$\alpha=0$, the radius $a(\tau)$ is a linear function of $\tau$.
Let me remind you that the above solutions are valid only in the
sector of small $\rho$. If $R_0$ is big enough, this does not
influence the case $\alpha >0$, but definitely excludes the vicinity
of singular points in the case $\alpha <0$.

At the end of this section, I want to draw the reader's attention to
the fact that the {\it obtained vacuum solution violates the equality
of all directions} inherited from the Euclidean bulk. Indeed, the
geometry (\ref{9}), with $R(\eta)$ given by (\ref{11}), is homogeneous
and isotropic in three compact directions, but not in the $\eta$
direction. As a consequence, the world-sheet is endowed with a
privileged direction. This is exactly the origin of the effective
Lorenzian metric derived in the next section. The asymmetry of the
vacuum solution itself, on the other hand, is a consequence of the
assumed compactness of the $3$-brane. In such a case, the brane must
be infinite in the remaining direction. Indeed, it is straightforward
to verify that the conservation equation (\ref{1}) forbids fully
compact world-sheets. In simple terms, matter can not be created out
of nothing. Therefore, the equality of all directions on the
world-sheet is violated already by the fact that the world-sheet has
only one infinite dimension.

\section{Stability analysis}\label{Sec3}

In this section, I shall examine the stability of the solution
(\ref{11}) against small perturbations of the brane. Without loss of
generality, the brane perturbations are defined by
$$
\delta x^{\mu} = \epsilon\, n^{\mu} ,
$$
where $n^{\mu}$ is the unit normal to the ground state wold-sheet, and
$\epsilon=\epsilon (\xi)$. This way, we are left with only one
component of the brane perturbations. This is a consequence of the
fact that 4-dimensional world-sheet $x^{\mu}(\xi)$ has only one
orthogonal direction in the considered 5-dimensional bulk. The other
variable is the variation of the brane stress tensor $\delta m^{ab}$.
In the model independent approach of this paper, $\delta m^{ab}$ is
not fully specified.

Let me start with rewriting the world-sheet equations (\ref{3}) in
the equivalent form
$$
m^{ab}K^{\mu}_{ab}=0 \,, \qquad \nabla_a m^{ab}=0 \,,
$$
where $K^{\mu}_{ab}$ stands for the extrinsic curvature of the brane
world-sheet,
$$
K^{\mu}_{ab}\equiv \nabla_a u^{\mu}_b \,.
$$
It is straightforwardly checked that the bulk vectors $K^{\mu}_{ab}$ are
orthogonal to the world-sheet. In the 5-dimensional bulk considered in
this paper, this implies the decomposition
$$
K^{\mu}_{ab} = n^{\mu} K_{ab}  \,.
$$
In terms of the extrinsic curvature $K^{\mu}_{ab}$ and the world-sheet
connection $\Gamma^a{}_{bc}$, the perturbed world-sheet equations
read:
$$
K^{\mu}_{ab}\, \delta m^{ab} +
m^{ab}\delta K^{\mu}_{ab} = {\cal O}_2 \,,
$$
$$
\nabla_a \delta m^{ab} + m^{cb}\delta\Gamma^a{}_{ca} +
m^{ac}\delta\Gamma^b{}_{ca} = {\cal O}_2 \,.
$$
The variation $\delta\Gamma^a{}_{bc}$ is easily calculated once the
variation of the induced metric is known. The needed expressions for
$\delta\gamma_{ab}$ and $\delta K^{\mu}_{ab}$ are obtained
straightforwardly. They read:
$$
\delta\gamma_{ab} = -2 K_{ab}\epsilon \,,
$$
$$
\delta K^{\mu}_{ab} =
n^{\mu}\left( \nabla_a\nabla_b - K^c{}_a K_{cb}\right)\epsilon -
u^{c\mu}K_{ab}\nabla_c\epsilon \,.
$$
With this, the perturbed world-sheet equations take the form
\begin{equation} \label{12}
m^{ab}\left( \nabla_a\nabla_b - K^c{}_a K_{cb}\right)\epsilon +
K_{ab}\,\delta m^{ab} = {\cal O}_2 \,,
\end{equation}
\begin{equation} \label{13}
\begin{array}{rcl}
&&\ds\left( m^{ac}\,\nabla^b K_{ac} +
m^{bc}\,\nabla_c K^a{}_a\right) \epsilon +{}     \\
&&\ds \left( 2m^{ac}K^b{}_a + m^{bc}K^a{}_a\right)
\nabla_c\epsilon - \nabla_a\,\delta m^{ab} = {\cal O}_2 \,.
\end{array}
\end{equation}
In the linear approximation I work with, the tensors $m^{ab}$ and
$K_{ab}$ take their vacuum values, and are straightforwardly calculated.
Precisely, the vacuum form of $m^{ab}$ is given by (\ref{6}),
(\ref{7}) and (\ref{10}) of the preceding section, while nonzero
components of $K_{ab}$ read:
$$
K_{00} = 3\alpha\,R\left(\frac{R_0}{R}\right)^{3\alpha}, \quad
K_{\alpha\beta}=-\frac{1}{R}\left(\frac{R_0}{R}\right)^{3\alpha}
\gamma_{\alpha\beta} \,.
$$
With given $m^{ab}$ and $K_{ab}$, one still lacks enough information
to solve the perturbation equations. Indeed, the form of the
variation $\delta m^{ab}$ remains unknown in the model independent
approach of the present work. The needed information is hidden in the
structure of the brane constituent matter.

In what follows, I shall restrain from specifying the model
Lagrangian of the world-sheet matter. Instead, I shall only assume
that {\it world-sheet matter does not couple to metric derivatives}.
Among matter fields that satisfy this assumption is commonly used
scalar field, but also electromagnetic and Yang-Mills fields. With
these constituent fields, the term $\delta m^{ab}$ lacks $\epsilon$
derivatives. Thus, the kinetic term of the equation (\ref{12}) is not
influenced by $\delta m^{ab}$.

The structure of the kinetic term is best seen if the dependence on
$\epsilon$ derivatives is explicitly shown. This is achieved by
rewriting the equation (\ref{12}) in the form
$$
{\arraycolsep 0em
\begin{array}{rcl}
\ds \frac{\rho}{R^2}\bigg[ && 
\ds \ddot\epsilon+\alpha\triangle\epsilon
+ \left(3\alpha - 1\right)\frac{\dot R}{R}\,\dot\epsilon         \\
&&\ds  {}-\alpha\left(\alpha + \frac{1}{3}\right) R^2
\left(K^{\alpha}_{\alpha}\right)^2\epsilon \bigg] +
K_{ab}\, \delta m^{ab} = {\cal O}_2 \,,
\end{array}}
$$
where $\triangle$ stands for the Laplacian of the unit $3$-sphere. To
simplify this expression, I introduce the auxiliary metric $\tilde
\gamma_{ab}$ defined by
\begin{equation}\label{14}
d\tilde s^2 \equiv - A \left\{ d\eta^2 + \frac{1}{\alpha}
\Big[ d\psi^2 + \sin^2\psi \left( d\theta^2 + \sin^2\theta\,
d\phi^2 \right)\Big]\right\}. 
\end{equation}
The shorthand notation
$$
A \equiv R_0^2 \left(\frac{R}{R_0}\right)^{3\alpha-1}
$$
is introduced for convenience.
Using this auxiliary metric, one can construct the box operator
$\tilde\Box \equiv \tilde\gamma^{ab}\, \tilde\nabla_a \tilde\nabla_b$.
It is checked that its action on the scalar $\epsilon(\xi)$ reduces to
$$
\tilde\Box\,\epsilon \equiv -\frac{1}{A} \bigg[\ddot\epsilon
+\alpha\triangle\epsilon
+\left(3\alpha - 1\right)\frac{\dot R}{R}\,\dot\epsilon \bigg] .
$$
Now, it is seen that the kinetic term of the above word-sheet equation
has the form $\tilde\Box\,\epsilon$. The complete
perturbation equation (\ref{12}) is rewritten as
\begin{equation} \label{15}
\left[\tilde\Box+3\alpha\left(3\alpha+1\right){B}\right]\epsilon -
{C} K_{ab}\,\delta m^{ab} = {\cal O}_2 \,,
\end{equation}
where the notation
$$
{B} \equiv \frac{1}{R^2_0}
\left(\frac{R_0}{R}\right)^{9\alpha-1}, \quad
{C} \equiv \frac{1}{\rho_0}\left(\frac{R_0}{R}\right)^{6(\alpha-1)}
$$
is introduced for convenience. The unknown variation $\delta m^{ab}$
can contribute to the mass and source terms of the equation
(\ref{15}), but it can not modify the kinetic term $\tilde \Box
\epsilon$. Therefore, the auxiliary metric $\tilde \gamma_{ab}$ must
have Minkowski signature to prevent the instability of the vacuum
solution $\epsilon = 0$. It follows then that vacuum stability
requires negative $\alpha$.
\begin{itemize}
\item {\it Vacuum stability requires $\alpha<0$}.
\end{itemize}
It should be noted that this is just a {\it necessary condition}, as
the full stability also requires a proper sign of the mass term,
which can not be fully determined without the knowledge of $\delta
m^{ab}$. Besides, the equation (\ref{15}) can also have a source. Its
origin is the brane matter, whose dynamics is only partially
determined by the equation (\ref{13}). For the full stability
analysis, one needs a model Lagrangian of the world-sheet matter. In
Sec. \ref{Sec5}, a specific example of the brane constituent matter
is considered in detail. In particular, the brane made of an
incompressible perfect fluid is shown to have stable dynamics
whenever $\alpha\leq -1/3$.

Let me conclude this section with the important observation that the
auxiliary metric $\tilde \gamma_{ab}$ is, in fact, the metric which
is detected in physical measurements. Indeed, the spacetime metric
is always determined by measuring behavior of test matter in curved
geometry. In my geometric approach, the {\it observable matter is
identified with brane vibrations}. As a consequence, the scalar
$\epsilon(\xi)$ is the only detectable matter in the case under
consideration. Its dynamics is governed by the Klein-Gordon type
equation (\ref{15}). This means that the observer that detects $\epsilon$
experiences $\tilde \gamma_{ab}$ as physical metric.
\begin{itemize}
\item {\it Auxiliary metric $\tilde \gamma_{ab}$ is seen as
physical spacetime metric}.
\end{itemize}
It should be noted that this result follows from a model which
provides only one scalar field to probe braneworld geometry. To be
realistic, the observable geometry must be probed with all the known
physical fields. As explained in the introduction, the geometric
approach of this work postulates brane vibrations as the only
observable matter. This means that the realistic braneworld must live
in a $D$-dimensional bulk characterized by $D\gg 5$. Only then the
variety of brane vibrations would be rich enough to represent all the
known physical fields. In a situation like this, the world-sheet
equations accommodate $D-4$ scalar fields. It is a difficult task to
find how conventional matter fields are connected to the brane
excitations. The simple model considered in this paper serves as a
guide of how realistic metric is constructed. Ultimately, one hopes
that all the conventional matter fields can be expressed in terms of
the brane excitations.

At the end of this section, I want to emphasize once more that brane
constituent matter is considered undetectable in our approach. One
can say that it parallels the dark energy of the conventional
cosmology. Basically, it means that excitations of the internal
structure of the brane are highly suppressed. With this, the
observable matter stems from the brane vibrations alone.

\section{Effective Universe}\label{Sec4}

As explained in the preceding section, the metric $\tilde\gamma_{ab}$
is the one that is detected in astronomical observations. For that
reason, I shall refer to it as the {\it effective metric} of the {\it
effective Universe}. In what follows, I shall restrict to the stable
sector $\alpha<0$. In this sector, the effective metric $\tilde
\gamma_{ab}$ has the needed Minkowski signature.

What are the properties of the effective Universe? To show this, I
shall first rewrite the solution (\ref{11}) in terms of the {\it
effective proper time} $\tilde\tau$, which is defined by the demand
that the effective interval takes the form
\begin{equation}\label{16}
d\tilde s^2 = -d\tilde\tau^2 + \tilde a^2 \left[ d\psi^2 +
\sin^2\psi \left( d\theta^2 + \sin^2\theta\, d\phi^2 \right) \right].
\end{equation}
The needed coordinate transformation reads
\begin{equation}\label{17}
\tilde\tau(\eta) = R_0 \int_0^{\eta}
\big[\cosh(3\alpha\eta)\big]^{\frac{3\alpha-1}{6\alpha}} d\eta \,.
\end{equation}
In terms of $\tilde\tau$, the effective radius $\tilde a$ becomes
\begin{equation} \label{18}
\tilde a \equiv \frac{R_0}{\sqrt{-\alpha}}
\left(\frac{R}{R_0}\right)^\frac{3\alpha-1}{2}.
\end{equation}
Thus, the {\it effective Universe is of bouncing type} for all
negative values of $\alpha$. It should be noted that the genuine
Universe considered in this section is, in fact, the Big Bang
Universe of Fig. \ref{f1}. It is the standard measuring procedure
that makes it look like a bouncing Universe. In other words, {\it
what we see is not what it actually is}.

Now, let me make use of the known experimental data to determine the
free parameters of the effective Universe. I start with the known
Hubble and deceleration parameters, whose definitions are given by
$$
H \equiv \frac{\dot{\tilde a}}{\tilde a}\,,\qquad
q \equiv -\frac{\ddot{\tilde a}}{\tilde a H^2}\,.
$$
(The {\it dot} denotes differentiation with respect to the effective
proper time: $\dot {\tilde a} \equiv d\tilde a /d\tilde\tau$.) In the
natural units $c=\hbar=1$, the reported present time values of these
two parameters are
$$
H_{\rm now} \approx 0.8 \times 10^{-26} \ m^{-1} \,, \qquad
q_{\rm now} \approx -0.5 \,.
$$
It should be noted, however, that these data refer to the realistic
Universe, characterized by the presence of both, dark energy and
observable matter. In the braneworld language, this means that, along
with the unavoidable brane constituent matter, the brane vibrations
should also be considered. This is not the case with the effective
Universe defined by (\ref{16}) and (\ref{18}). Indeed, this solution
refers to the unperturbed ground state of the Universe. Nevertheless,
these formulas can be used for the description of a realistic
Universe, too. This is because the assumption of strict spatial
homogeneity and isotropy can be replaced by the equally acceptable
{\it approximate homogeneity and isotropy}. In this picture, the
brane vibrations are uniformly distributed over the brane, so that
their contribution to the stress tensor $m^{ab}$ does not violate its
vacuum form (\ref{6}). Thus, the formulas (\ref{16}) and (\ref{18})
are applicable to the realistic cosmology, provided $\rho$ and $p$
are understood as spatially averaged quantities that define cosmic
matter.

With this, I am able to determine the present time of the Universe
$\tilde\tau_{\rm now}$, and the value of the constant $R_0$. The
result is shown in Table \ref{t1}.
\begin{table}
\caption{Present time parameters of the Universe. \label{t1}}
\vspace*{3mm}
\begin{ruledtabular}
\begin{tabular}{cccc}
$\alpha$ & $R_0$ & $\eta_{\rm now}$ & $\tilde\tau_{\rm now}$ \\
\hline
$-0.01$  &  $0.8\times 10^{25}\,$ m & 11.2 & $13.7\,$ Gyr   \\
$-0.1$   &  $2.8\times 10^{25}\,$ m & 2.84 & $10.9\,$ Gyr   \\
$-1$     &  $1.4\times 10^{26}\,$ m & 0.44 & $7.6 \,$ Gyr    \\
$-10$    &  $1.1\times 10^{27}\,$ m & 0.05 & $6.8 \,$ Gyr    \\
\end{tabular}
\end{ruledtabular}
\end{table}
It should be noted that the obtained results are as approximate as
the experimental values of $H_{\rm now}$ and $q_{\rm now}$ are. For
example, the measured value of $q_{\rm now}$ is allowed to take
values in the interval $-0.63 \lesssim q_{\rm now} \lesssim -0.36$
\cite{SCA}. As a consequence, the experimental error in determining
the age of the Universe may exceed $1$ Gyr. Another interesting
observation is that the present time of the Universe falls into the
interval
$$
6.7\ {\rm Gyr} \lesssim \tilde\tau_{\rm now} \lesssim 14.2\ {\rm Gyr}
$$
whatever negative value of the parameter $\alpha$ is taken. Indeed,
the direct calculation shows that $\tilde\tau_{\rm now} \to 14.2$ Gyr
in the limit $\alpha\to 0$, while $\alpha\to -\infty$ implies
$\tilde\tau_{\rm now} \to 6.7$ Gyr.

In what follows, I shall rewrite my results in the form suitable for
the comparison with the standard GR cosmology. To this end, the
effective metric $\tilde\gamma_{ab}$ is used for the calculation of
the effective curvature, and the construction of the effective
Einstein's equations. As a result, one obtains
\begin{equation} \label{19}
\tilde{\cal R}_{ab}-\frac{1}{2}\,\tilde\gamma_{ab}\,\tilde{\cal R}=
\kappa \,\tilde m_{ab} \,,
\end{equation}
where $\tilde m_{ab}$ stands for the effective stress-energy, and
$\kappa$ is identified with the gravitational constant. The direct
calculation yields the stress-energy of the homogeneous and isotropic
perfect fluid
\begin{equation} \label{20}
\tilde m^{ab} = \tilde p \,\tilde\gamma^{ab} -
\left(\tilde\rho + \tilde p\right)\tilde\gamma^{a0}\delta^b_0 \,,
\end{equation}
where the effective energy density and pressure are given by the
following expressions:
\begin{subequations} \label{21}
\begin{equation}\label{21a}
{\arraycolsep 0em
\begin{array}{rcl}
\ds \tilde\rho \equiv \frac{\tilde\rho_0}{4\alpha}
\Bigg[ && \ds \left(1-\alpha\right) \left(9\alpha-1\right)
\left(\frac{R_0}{R}\right)^{3\alpha-1}                    \\  
&& \ds {}+\left(3\alpha-1\right)\left(3\alpha-1\right)
\left(\frac{R_0}{R}\right)^{9\alpha-1}\Bigg],          \\ 
\end{array}}
\end{equation}
\begin{equation}\label{21b}
{\arraycolsep 0em
\begin{array}{rcl}
\ds \tilde p \equiv \frac{\tilde\rho_0}{12\alpha}
\Bigg[ && \ds \left(1-\alpha\right) \left(1-9\alpha\right)
\left(\frac{R_0}{R}\right)^{3\alpha-1}                       \\ 
&& \ds {}+\left(3\alpha-1\right)\left(9\alpha+1\right)
\left(\frac{R_0}{R}\right)^{9\alpha-1}\Bigg].             \\        
\end{array}}
\end{equation}
\end{subequations}
Here, $\tilde\rho_0\equiv \tilde\rho(0)$, and the coupling constant
$\kappa$ is defined as
\begin{equation} \label{22}
\kappa \equiv -\frac{3\alpha}{\tilde\rho_0 R^2_0} \,.
\end{equation}
Using data from Table \ref{t1}, and the known value of the
gravitational constant,
$$
\kappa = 6.4 \times 10^{-69}\ {\rm m}^2 ,
$$
one is able to calculate the constant $\tilde\rho_0$ as a function of
$\alpha$. Then, the equations (\ref{21}) yield the present values of
the effective energy density and pressure, as well as the effective
equation of state $\tilde p =\tilde p(\tilde\rho)$. Locally, it can
be written in the familiar form $\tilde p = \tilde\alpha \,
\tilde\rho$. The parameters of the effective GR cosmology at the
present epoch are displayed in Table \ref{t2}.
\begin{table}
\caption{Present epoch of the effective GR cosmology.
\label{t2}}
\vspace*{3mm}
\begin{ruledtabular}
\begin{tabular}{ccccc}
$\alpha$ & $\tilde\alpha_{\rm now}$ & $\tilde\rho_0$ &
$\tilde\rho_{\rm now}$ & $\tilde\Omega_{\rm now}$   \\
\hline
$-0.01$ & $-0.58$ & $7.3\times 10^{16}\,{\rm m}^{-4}$ &
$4.2\times 10^{16}\,{\rm m}^{-4}$  & 1.4     \\
$-0.1$  &  $-0.56$ & $6.0\times 10^{16}\,{\rm m}^{-4}$ &
$4.4\times 10^{16}\,{\rm m}^{-4}$  & 1.5   \\
$-1$     &  $-0.58$ & $2.4\times 10^{16}\,{\rm m}^{-4}$ &
$3.8\times 10^{16}\,{\rm m}^{-4}$  & 1.3   \\
$-10$   &  $-0.65$ & $0.4\times 10^{16}\,{\rm m}^{-4}$ &
$3.5\times 10^{16}\,{\rm m}^{-4}$  & 1.2   \\
\end{tabular}
\end{ruledtabular}
\end{table}
The density parameter $\tilde\Omega$, displayed in the last column,
is defined by $\tilde\Omega\equiv\kappa\tilde\rho/3H^2$.

A few remarks are in order. First, the data displayed in Table
\ref{t2} refer to the total matter content of the effective Universe.
In particular, the effective density parameter $\tilde\Omega$ is what
is conventionally denoted by $\Omega_{\rm tot}$. Second, the
effective GR cosmology considered in this section can be modified by
the inclusion of the cosmological constant. This means that the
equation (\ref{19}) can be rewritten as
\begin{equation} \label{23}
\tilde{\cal R}_{ab}-\frac{1}{2}\,\tilde\gamma_{ab}
\left( \tilde{\cal R} + \Lambda \right) = \kappa\,\hat m_{ab} \,,
\end{equation}
where $\hat m_{ab}$ is the modified stress energy, whose energy
density and pressure read:
\begin{equation} \label{24}
\hat\rho \equiv \tilde\rho + \frac{\Lambda}{2\kappa}\,,\qquad
\hat p \equiv \tilde p - \frac{\Lambda}{2\kappa}\,.
\end{equation}
Now, one can calculate the modified values of the cosmological
parameters. First, by the inspection of Table \ref{t1}, one finds
that the choice $\alpha=-0.01$ gives the best fit for the measured
age of the Universe. Then, the cosmological constant $\Lambda$ is
determined by using (\ref{24}) to enforce the reported value of the
total density parameter $\hat\Omega_{\rm now} \approx 1$. As a
result,
$$
\Lambda \approx -1.5 \times 10^{-52}\, {\rm m}^{-2} .
$$
With this value of $\Lambda$, the calculated cosmological parameters
become close to those commonly reported. In particular,
$$
\hat\tau_{\rm now} \approx 13.7\ {\rm Gyr}\,,\qquad
\hat\Omega_{\rm now}\approx 1\,.
$$
It should be noted that the notion of the age of the Universe in this
paper is not what is commonly used in the standard cosmology. This is
because the effective Universe (\ref{18}) has no beginning. As a
substitute for the age of the Universe, I use the notion of the
present time as measured from the bounce. I also remind you that all
my calculations refer to the sector of small $\rho$. As a consequence,
the obtained results do not cover the whole history of the Universe,
and the cosmological parameters such as the age of the Universe can
not be reliably determined. In fact, I shall demonstrate in the next
section that typical braneworld concept does not support Universes of
finite lifetime.

\section{Example}\label{Sec5}

In this section, I shall analyze a brane made of an {\it incompressible
perfect fluid} with the equation of state $p=\alpha\rho$. It should be
emphasized that this is just a model for the brane stress tensor
$m^{ab}$. In a more realistic approach, the behavior of the brane
constituent matter is derived from a bulk Lagrangian.

The stress tensor of the world-sheet perfect fluid has the form
\begin{equation} \label{25}
m^{ab} = p\gamma^{ab} + \left(\rho-p\right) v^a v^b \,,
\end{equation}
where $v^a\equiv d\xi^a/ds$ is the covariant $4$-velocity of the
fluid. The $4$-velocity is subject to the constraint $\gamma_{ab}\,v^a
v^b=1$, and $\rho$ and $p$ are connected by the equation of state
(\ref{7}). When spatial distribution of the fluid is maximally
symmetric, the stress tensor (\ref{25}) takes the homogeneous and
isotropic form (\ref{6}).

Now, one has enough information to solve the perturbation equations
(\ref{12}) and (\ref{13}). As it turns out, the first equation
decouples from the second, and takes the form
\begin{equation} \label{26}
\tilde\Box\,\epsilon - 3\alpha\left(3\alpha+1\right) \frac{1}{R_0^2}
\left(\frac{R_0}{R}\right)^{9\alpha-1}\epsilon\, = \,{\cal O}_2 \,.
\end{equation}
The stability of the Klein-Gordon-like equation (\ref{26}) requires
Minkowski signature of its kinetic term, and negative mass term. As a
consequence, the necessary and sufficient condition for the stable
brane perturbations becomes
$$
\alpha \leq -\frac{1}{3} \,.
$$
The second equation, on the other hand, yields unstable behavior for
the independent variations $\delta\rho$ and $\delta \vec v$. As it
turns out, the general solution of (\ref{13}) takes the form
\begin{subequations} \label{27}
\begin{equation} \label{27a}
\alpha\left(\frac{\delta\rho}{\rho} - K^a_a\,\epsilon\right) =
\left(\alpha-1\right) \left(\frac{R_0}{R}\right)^{3\alpha+1}
\dot\omega + {\cal O}_2 \,,
\end{equation}
\begin{equation} \label{27b}
R\,\delta\vec v = \left(\frac{R_0}{R}\right)^{3\alpha+1}
\left(\vec\omega + \vec\nabla\omega\right) + {\cal O}_2 \,,
\end{equation}
\end{subequations}
where the small residual coefficients $\vec\omega(\xi)$ and
$\omega(\xi)$ are constrained by
\begin{subequations} \label{28}
\begin{equation} \label{28a}
\vec\nabla \cdot \vec\omega = \dot{\vec\omega} = 0 \,,
\end{equation}
\begin{equation} \label{28b}
\ddot\omega - \alpha\,\triangle\omega - \left(3\alpha+1\right)
\frac{\dot R}{R}\,\dot\omega = 0 \,.
\end{equation}
\end{subequations}
Here, $\vec\nabla$ stands for the $3$-dimensional covariant
derivative on the unit $3$-sphere, $\triangle$ is the corresponding
Laplacian, and the "dot" denotes differentiation with respect to
$\eta$. As seen from (\ref{28b}), the condition $\alpha<0$ makes the
coefficient $\omega$ unstable. Thus, the {\it stability of the brane
vibrations is incompatible with the stability of the brane interior}.

Having in mind the assumed generality of the brane fluid, the
obtained result is not that surprising. Indeed, it is difficult to
imagine that our braneworld is made of, let us say, a rare gas. In
what follows, I shall pursue a more plausible idea that the brane
constituent fluid is {\it incompressible}. This is achieved by
imposing the additional constraint
\begin{equation} \label{29}
\vec\nabla \cdot \delta\vec v = 0 \,.
\end{equation}
With (\ref{29}), the equation (\ref{27b}) reduces to $\triangle \omega
= 0$, which implies $\omega = \omega(\eta)$. Indeed, among eigen
values of the Laplacian on the $3$-sphere there is no zero eigen
value. This leaves us with a spatial constant as the only solution for
$\omega$. With the absence of spatially localized perturbations, the
general solution $\omega = \omega(\eta)$ is physically equivalent to
$\omega = 0$. Then, the equations (\ref{27}) reduce to
\begin{subequations} \label{30}
\begin{equation} \label{30a}
\frac{\delta\rho}{\rho} = K^a_a\,\epsilon + {\cal O}_2 \,,
\end{equation}
\begin{equation} \label{30b}
R\,\delta\vec v = \left(\frac{R_0}{R}\right)^{3\alpha+1}
\vec\omega + {\cal O}_2 \,.
\end{equation}
\end{subequations}
The first equation is a constraint equation, as $\delta\rho$ has no
degrees of freedom. The second equation tells us that fluid flow has
only rotational degrees of freedom. Owing to the constraint
(\ref{28a}), these degrees of freedom are effectively frozen. Indeed,
the equations (\ref{28a}) and (\ref{30b}) ensure that initially small
fluid flow remains to be small for all $\eta$. Thus, the {\it brane
interior is stable against its small perturbations}.

The comparison with the perfect fluid brane in Minkowski bulk is
straightforward. It is easily found that the latter can never be
stable. Indeed, the correct kinetic term requires $\alpha>0$, whereas
the correct mass term is obtained only if $-1/3\leq\alpha\leq 0$. This
is exactly the motivation for considering Euclidean bulks.

\begin{figure}[htb]
\begin{center}
\includegraphics[height=5cm]{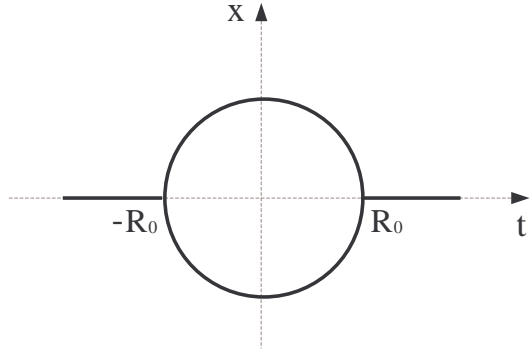}
\end{center}
\vspace*{-.5cm}
\caption{Creation of maximally symmetric Universe.\label{f2} }
\end{figure}

At the end of this subsection, let me show how the actual $4$-surface
that represents our braneworld looks in the $5$-dimensional Euclidean
bulk. While its parametric form is given by (\ref{4}), its implicit
form reads
$$
(x^1)^2 + (x^2)^2 + (x^3)^2 + (x^4)^2 = \ell^2(t) \,.
$$
To simplify exposition, I shall consider the simplest case $\alpha =
-1/3$. Then, the radius $\ell(t)$ is found to have the simple form
$$
\ell^2 = R^2_0 - t^2 \,.
$$
With this, the sought $4$-surface takes the form of a regular
$4$-sphere in the $5$-dimensional Euclidean bulk:
\begin{equation} \label{31}
(x^0)^2 + (x^1)^2 + (x^2)^2 + (x^3)^2 + (x^4)^2 = R^2_0 \,.
\end{equation}
There are two reasons why this solution looks odd. The first stems
from the fact that the geometry of the surface (\ref{31}) is
maximally symmetric, and therefore, has no privileged direction. How
is it possible then that the behavior of brane vibrations defines
effective time? The answer is, in fact, very simple. The brane
world-sheet is characterized not only by its geometry, but also by
the distribution of matter. A simple calculation yields
$$
\rho = \rho_0 \left(\frac{R^2_0}{R^2_0 - t^2}\right)^2 ,
$$
which reveals the anisotropy of $\rho(\xi)$. Thus, it is the gradient
of $\rho$ which determines the effective time direction. The second
reason to doubt the solution (\ref{31}) comes from its
incompatibility with the conservation equation (\ref{1}). Indeed, the
surface (\ref{31}) is a finite $4$-sphere located in the area
$-R_0\leq t\leq R_0$. This means that the brane did not exist before
$t=-R_0$ when it was created {\it out of nothing}. The explanation of
this paradox lies in the fact that the solution under consideration
is obtained by cutting the $t$-axis at singular points $t=\pm R_0$.
One can cure this by employing the trivial solution $\ell(t)=0$ on
the rest of the $t$-axes. The obtained modification is depicted in
Fig. \ref{f2}.
It shows an ever existing massive point particle that explodes at
$t=-R_0$, expands to its maximum at $t=0$, and then shrinks back to a
point at $t=R_0$. This solution is fully compatible with the bulk
conservation equation (\ref{1}). An important conclusion is that the
bulk conservation equations do not allow fully bounded braneworld
cosmologies. As typical bulk matter respects conventional
conservation laws, a typical braneworld can not be created out of
nothing.
\begin{itemize}
\item {\it Typical braneworld concept does not support Big Bang
cosmologies}.
\end{itemize}
To be more precise, the above statement holds in geodesically
complete bulks of trivial topology, irrespectively of the bulk
signature.

\section{Recapitulation}\label{Sec6}

The purpose of this work has been to demonstrate how a stable
conventional Universe emerges from a brane embedded in Euclidean
bulk. To this end, I had to solve two major problems. First, the
stability of the brane dynamics is ensured by employing Euclidean,
rather than Minkowskian bulk. Second, I demonstrated how a particular
geometry of Euclidean signature can effectively be seen as
Minkowskian. To achieve this, I have examined the behavior of a
$3$-brane in a $5$-dimensional Euclidean bulk. No bulk action was
initially specified. Instead, the brane equations are derived from
the universally valid conservation equations of the bulk stres
tensor. Specifically, I have investigated small perturbations of the
$3$-sphere. The corresponding world-sheet is interpreted as the
Universe.

The obtained results are summarized as follows. First, the equation
of state $p=\alpha\rho$ for maximally symmetric $3$-branes has been
shown to have a universal character in the sector of small $\rho$.
This way, the general form of the unperturbed brane equations has
been derived, irrespectively of the nature of the brane constituent
matter. Second, the stability analysis has lead me to the constraint
$\alpha<0$, as a necessary condition for the stable brane world-sheet
in Euclidean bulk. The corresponding perturbation equation turns out
to have a Klein-Gordon-like kinetic term, based on the auxiliary
metric of Minkowski signature. It has been argued that it is this
auxiliary metric that is detected in physical measurements. Indeed,
the spacetime geometry is always probed by some kind of test matter.
In our case, the observed matter stems from the brane vibrations. As
these are guided by a Klein-Gordon-like equation, their detection
reveals the above mentioned auxiliary metric as a physical metric of
the Universe. The third result concerns the effective Universe thus
obtained. Its unperturbed evolution has been rewritten in the
conventional GR form, and compared with the standard cosmology. The
present epoch values of the average energy density and pressure are
calculated along with the effective cosmological constant and the age
of the Universe. Unfortunately, the more complex issues, such as
creation of structure or the origin of anisotropies, cannot be
addressed in this simple model. Indeed, the very inclusion of matter
requires a bulk of dimension $D\gg 5$. Finally, I have demonstrated
that braneworld concept rules out the Big Bang cosmology. This is a
consequence of the assumed bulk conservation equations, which are
incompatible with the creation out of nothing.

At the end, let me explain the philosophy that lies behind the main
achievement of this paper. Basically, it is the idea to obtain the
observable physics as a physics of brane vibrations. This is a kind
of geometrization program which is not easy to achieve. The main
obstacle stems from the fact that brane constituent matter has its
own dynamics. To construct a plausible theory of brane vibrations,
one should find a mechanism to suppress internal brane dynamics. The
problem is similar to the problem of unobservable extra dimensions in
Kaluza-Klein theories. There, the extreme smallness of the internal
space solves the problem. In the braneworld approach, one could
require an extreme brane tension. The search for a bulk Lagrangian
that supports a high tension brane is a difficult task, which will be
addressed elsewhere.

\begin{acknowledgments}
This work is supported by the Serbian Ministry of Education, Science
and Technological Development, under Contract No. $171031$.
\end{acknowledgments}

\end{document}